\documentclass[12pt]{article}
\usepackage{bbm}
\newcommand{\ppsl}{p \hskip-5pt /}
\newcommand{\kksl}{k \hskip-6pt /}
\newcommand{\dloop}{\delta^\mathrm{1\mbox{-}loop}}
\begin{document}
\title{\normalsize \hfill UWThPh-2002-21 \\[1cm] \LARGE
One-loop corrections to the seesaw mechanism \\
in the multi-Higgs-doublet Standard Model}
\author{
Walter Grimus\thanks{E-mail: grimus@doppler.thp.univie.ac.at} \\
\small Institut f\"ur Theoretische Physik, Universit\"at Wien \\
\small Boltzmanngasse 5, A--1090 Wien, Austria
\\*[3.6mm] \addtocounter{footnote}{2}
Lu\'{\i}s Lavoura\thanks{E-mail: balio@cfif.ist.utl.pt} \\
\small Universidade T\'ecnica de Lisboa \\
\small Centro de F\'\i sica das Interac\c c\~oes Fundamentais \\
\small Instituto Superior T\'ecnico, P--1049-001 Lisboa, Portugal
\\*[4.6mm]
}

\date{11 September 2002}
\maketitle

\begin{abstract}
We consider the lepton sector of the Standard Model
and allow for an arbitrary number of Higgs doublets and,
moreover,
for the presence of right-handed neutrino singlets
which enable the seesaw mechanism.
In this framework,
we identify and calculate the dominant one-loop radiative corrections
to the tree-level mass matrix of the light neutrinos.
The interesting feature is that
both the tree-level and the one-loop contributions
to the light-neutrino mass matrix are quadratic in the Yukawa couplings,
with the effect that
the one-loop contribution is smaller than the tree-level one
mainly because of the one-loop factor $\left( 16\pi^2 \right)^{-1}$.
We also point out the possibility
of generating radiatively---in this framework---the ratio
of solar over atmospheric neutrino mass-squared differences,
as needed for the large-mixing-angle MSW solution
of the solar-neutrino problem.

\vspace*{8mm}

\normalsize\noindent
PACS numbers: 14.60.Pq, 14.60.St
\end{abstract}

\newpage

\section{Introduction}

It is believed that the final confirmation
of small but non-zero neutrino masses is around the corner,
since neutrino oscillations \cite{pontecorvo}
provide an excellent and natural solution
for the atmospheric- and solar-neutrino problems.
In the latter case,
matter effects in the neutrino oscillations \cite{MSW}
seem to play a crucial role.
For reviews,
see for instance Ref.~\cite{reviews};
for the latest solar-neutrino results,
see Ref.~\cite{holanda} and the references therein. 

The results of the solar- and atmospheric-neutrino experiments
have brought about an upsurge of model building
for the lepton masses and mixings---for reviews
of lepton-mass-matrix textures and models see Ref.~\cite{models}.
In this effort,
the seesaw mechanism \cite{seesaw} figures prominently
as a means for obtaining small Majorana masses for the neutrinos.

In this paper we discuss a very simple extension
of the Standard Model (SM):
we add to the SM right-handed neutrino singlets,
and moreover admit an arbitrary number of Higgs doublets.
The first addition allows to incorporate the seesaw mechanism in the SM.
As shown in Ref.~\cite{GL01},
this simple extension of the SM
provides a framework in which interesting models can be constructed,
which reproduce in a natural way
at least part of the striking features of the neutrino masses and mixings.
This extension of the SM has two large mass scales:
the electroweak scale,
of about 100 GeV,
which is the order of magnitude
of the $Z$ and $W$ boson masses $m_Z$ and $m_W$,
respectively,
and also of the scalar masses;
and the seesaw scale $m_R$,
much larger than $m_Z$,
which is the scale of the singlet-neutrino masses. 
A small mass scale $m_D$,
below the electroweak scale,
gives the order of magnitude
of the elements of the Dirac neutrino mass matrix $M_D$,
in such a way that
the light-neutrino masses end up being of order $m_D^2 / m_R$. 

It was noticed in Ref.~\cite{GN89} that,
in this framework,
the dominant radiative corrections to the seesaw mechanism
are quadratic in the Yukawa couplings,
just like the tree-level masses,
and that the radiative corrections are smaller than the tree-level results
solely because of the appearance
of the factor $\left( 16\pi^2 \right)^{-1}$ in one-loop integrals.
By using two Higgs doublets and only one right-handed neutrino singlet,
this fact was later exploited in Ref.~\cite{GN00}
for the radiative generation of the solar mass-squared difference;
the ratio $\Delta m^2_\odot/\Delta m^2_\mathrm{atm}$ 
of solar over atmospheric mass-squared differences
comes out of the correct order of magnitude
for the large-mixing-angle MSW solution of the solar-neutrino problem
(for the main features of that solution,
with $\Delta m^2_\odot / \Delta m^2_\mathrm{atm}$ being about 0.02,
see for instance Ref.~\cite{holanda}),
with no other scale free to make adjustments. 

We want to extend here the calculations of Ref.~\cite{GN89},
where the focus was the radiative generation
of neutrino masses vanishing at tree level.
In the present work we perform
a general calculation of the dominant one-loop corrections
to the seesaw mechanism,
with emphasis on the demonstration of the finiteness and gauge-independence
(in $R_\xi$ gauges)
of those radiative corrections.
As a result of this effort we shall present a formula,
easily applicable in our extension of the SM,
whose purpose is twofold:
\begin{itemize}
\item To check the radiative stability of the seesaw mass matrix
and of the relations derived therefrom,
or else to compute the radiative corrections to those relations;
\item To explore the possibilities of radiatively generating 
$\Delta m^2_\odot/\Delta m^2_\mathrm{atm}$.
\end{itemize}
Unfortunately,
the second point appears quite difficult to us,
when one wants to combine it
with specific properties of the lepton mixing matrix $U$,
like the small $U_{e3}$ and the practically maximal atmospheric mixing.

This paper is organized as follows.
In Section~\ref{framework} we review the seesaw mechanism
and our formalism for the multi-Higgs-doublet SM.
We discuss the calculation of the neutrino self-energies
and write down their one-loop integrals in Section~\ref{self-energy}.
The actual calculation of the dominant one-loop corrections
to the seesaw mechanism 
and the demonstration of their finiteness and gauge-independence
are carried out in Section~\ref{deltaML}.
In Section~\ref{general} we give arguments in favour of the dominance
of the corrections calculated in the previous section.
In Section~\ref{concl} we present the final results and a summary.

\section{Framework}
\label{framework}

We consider the lepton sector of an extension of the Standard Model
with $n_L$ left-handed doublets and $n_L$ right-handed charged singlets,
i.e.,
$n_L$ families,
plus $n_R$ right-handed neutrino fields and $n_H$ Higgs doublets.
We denote the Higgs doublets by $\phi_k$
($k = 1, 2, \ldots, n_H$)
and define $\tilde \phi_k = i \tau_2 \phi_k^\ast$.
The vacuum expectation value (VEV) of the neutral component of $\phi_k$
is $v_k / \sqrt{2}$.
The Yukawa Lagrangian of the leptons is given by
\begin{equation}
\mathcal{L}_\mathrm{Y} = - \sum_{k=1}^{n_H}\, 
\left( \phi_k^\dagger \bar \ell_R \Gamma_k
+ \tilde \phi_k^\dagger \bar \nu_R \Delta_k \right) D_L
+ \mathrm{H.c.}
\label{Yukawa}
\end{equation}
We employ a vector and matrix notation where $\ell_R$,
$\nu_R$,
and $D_L$ are the vectors of the right-handed charged-lepton fields,
of the right-handed neutrino singlets,
and of the left-handed lepton doublets,
respectively.
The Yukawa coupling matrices $\Gamma_k$ are $n_L \times n_L$,
while the $\Delta_k$ are $n_R \times n_L$.
The charged-lepton mass matrix $M_\ell$
and the Dirac neutrino mass matrix $M_D$ are
\begin{equation}\label{Mell,D}
M_\ell = \frac{1}{\sqrt{2}}\, \sum_k v_k^\ast \Gamma_k
\quad {\rm and} \quad
M_D = \frac{1}{\sqrt{2}}\, \sum_k v_k \Delta_k\, ,
\end{equation}
respectively.
Without loss of generality,
we assume $M_\ell$ to be diagonal with real and positive diagonal elements:
$M_\ell = {\rm diag} \left( m_e, m_\mu, m_\tau, \ldots \right)$.
The mass terms for the neutrinos are
\begin{equation}\label{MR}
- \bar \nu_R M_D \nu_L
- \frac{1}{2}\, \bar \nu_R C M_R \bar \nu_R^T
+ {\rm H.c.}\, ,
\end{equation}
where $C$ is the charge-conjugation matrix
and $M_R$ is non-singular and symmetric. 
It is well known \cite{schechter}
that Eq.~(\ref{MR}) can be written in a compact form
as a mass term with an $(n_L+n_R) \times (n_L+n_R)$ symmetric mass matrix
\begin{equation}\label{D+M}
\mathcal{M}_{D+M} = 
\left( \begin{array}{cc} 0 & M_D^T \\
M_D & M_R \end{array} \right),
\end{equation}
which may be diagonalized as
\begin{equation}\label{Mtotal}
\mathcal{U}^T \mathcal{M}_{D+M}\, \mathcal{U} = \hat m
= \mathrm{diag} \left( m_1, m_2, \ldots, m_{n_L+n_R} \right),
\end{equation}
where the $m_i$ are real and non-negative.
In order to implement the seesaw mechanism \cite{seesaw}
we assume that the elements of $M_D$ are of order $m_D$
and those of $M_R$ are of order $m_R$,
with $m_D \ll m_R$.
Then,
the neutrino masses $m_i$ with $i=1, 2, \ldots, n_L$
are of order $m_D^2/m_R$,
while those with $i = n_L+1, \ldots, n_L+n_R$ are of order $m_R$. 
It is useful to decompose the $(n_L+n_R) \times (n_L+n_R)$ unitary
matrix $\mathcal{U}$ as \cite{GN89,GL02}
\begin{equation}\label{U}
\mathcal{U} = \left( \begin{array}{c} U_L \\ U_R^\ast \end{array} \right),
\end{equation}
where the submatrix $U_L$ is $n_L \times (n_L+n_R)$
and the submatrix $U_R$ is $n_R \times (n_L+n_R)$.
With these submatrices, 
the left- and right-handed neutrinos
are written as linear superpositions
of the $n_L+n_R$ physical Majorana neutrino fields $\chi_i$:
\begin{equation}\label{chi}
\nu_L = U_L \gamma_L \chi
\quad \mathrm{and} \quad
\nu_R = U_R \gamma_R \chi\, ,
\end{equation}
where $\gamma_L = \left( 1 - \gamma_5 \right) / 2$
and $\gamma_R = \left( 1 + \gamma_5 \right) / 2$
are the projectors of chirality.

The unitarity relations for $\mathcal{U}$ are \cite{GL02}
\begin{equation}\label{uni}
U_L U_L^\dagger = \mathbbm{1}_{n_L} \,, \quad
U_R U_R^\dagger = \mathbbm{1}_{n_R} \,, \quad
U_L U_R^T = 0_{n_L \times n_R} \,, \quad \mathrm{and} \quad
U_L^\dagger U_L + U_R^T U_R^* = \mathbbm{1}_{n_L+n_R} \,.
\end{equation}
From Eq.~(\ref{Mtotal}) we shall need the two relations
\begin{equation}\label{0}
U_L^* \hat m U_L^\dagger = 0 \quad \mathrm{and} \quad
U_R \hat m U_L^\dagger = M_D\,.
\end{equation}
It follows from Eqs.~(\ref{uni}) and (\ref{0}) that
\begin{equation}
U_R^\dagger M_D = U_R^\dagger U_R \hat m U_L^\dagger
= \left( \mathbbm{1}_{n_L+n_R} - U_L^T U_L^\ast \right) \hat m U_L^\dagger
= \hat m U_L^\dagger\, .
\label{relation}
\end{equation}

The leptonic charged-current Lagrangian is
\begin{equation}
\mathcal{L}_{\mathrm{cc}} = \frac{g}{\sqrt{2}}\, W_\mu^-
\bar \ell \gamma^\mu \gamma_L U_L \chi + \mathrm{H.c.}\, ,
\end{equation}
where $g$ is the $SU(2)$ gauge coupling constant.
The interaction of the $Z$ boson with the neutrinos is given by
\begin{equation}
\mathcal{L}_{\mathrm{nc}}^{(\nu)} = \frac{g}{4 c_w}\, Z_\mu
\bar \chi \gamma^\mu \left[ \gamma_L \left( U_L^\dagger U_L \right)
- \gamma_R \left( U_L^T U_L^\ast \right) \right] \chi\, ,
\end{equation}
where $c_w$ is the cosine of the Weinberg angle.

A full account of our formalism for the scalar sector
of the multi-Higgs-doublet SM is given in Ref.~\cite{GL02}.
We use the notation $S_b^0$ for the neutral-scalar mass eigenfields;
the vectors $b \in \mathbbm{C}^{n_H}$ characterize
each of those fields---for their precise definitions
see Refs.~\cite{GN89,GL02}. 
The Yukawa couplings of the neutral scalars $S^0_b$ 
to the neutrinos are given by
\begin{equation}
\mathcal{L}_\mathrm{Y}^{(\nu)} \left( S^0 \right) = - \frac{1}{2 \sqrt{2}}\,
\sum_b S^0_b\, \bar \chi \left[
\left( U_R^\dagger \Delta_b U_L
+ U_L^T \Delta_b^T U_R^\ast \right) \gamma_L
+ \left( U_L^\dagger \Delta_b^\dagger U_R
+ U_R^T \Delta_b^\ast U_L^\ast \right) \gamma_R
\right] \chi \,,
\label{neutralYuk}
\end{equation}
with $\Delta_b = \sum_k b_k \Delta_k$.
The neutral Goldstone boson $G^0 = S_{b_Z}^0$
is given by the vector $b_Z$ with
$\left( b_Z \right)_k = i v_k / v$ \cite{GN89,GL02}.
Here,
$v = \left( \left| v_1 \right|^2 + \left| v_2 \right|^2
+ \cdots + \left| v_{n_H} \right|^2 \right)^{1/2}
= 2 m_W / g$.
We thus obtain
\begin{equation}
\Delta_{b_Z} = \frac{ig}{\sqrt{2} m_W}\, M_D \,.
\label{utiwm}
\end{equation}
Similarly,
the charged-scalar mass eigenfields $S_a^\pm$
are characterized by vectors $a \in \mathbbm{C}^{n_H}$.
The Yukawa couplings of the charged scalars are given by
\begin{equation}
\mathcal{L}_{\mathrm{Y}} \left( S^\pm \right) =
\sum_a S_a^- \bar \ell \left[ \gamma_R \left( \Delta_a^\dagger U_R \right)
- \gamma_L \left( \Gamma_a U_L \right) \right] \chi
+ \mathrm{H.c.}\, ,
\end{equation}
where $\Delta_a = \sum_k a_k \Delta_k$ and
$\Gamma_a = \sum_k a_k^\ast \Gamma_k$.
The charged Goldstone boson $G^\pm = S_{a_W}^\pm$
is given by the vector $a_W$ with $\left( a_W \right)_k = v_k / v$;
therefore
\begin{equation}
\Delta_{a_W} = \frac{g}{\sqrt{2} m_W}\, M_D
\quad \mathrm{and} \quad
\Gamma_{a_W} = \frac{g}{\sqrt{2} m_W}\, M_\ell\, .
\label{ugjsl}
\end{equation}

\section{Neutrino self-energies}
\label{self-energy}

The seesaw mechanism \cite{seesaw} tells us that,
at tree level,
the mass matrix of the light neutrinos is given by
\begin{equation}\label{Mtree}
\mathcal{M}_\nu^\mathrm{tree} = -M_D^T M_R^{-1} M_D \,,
\end{equation}
where $M_D$ and $M_R$ appear in the full mass matrix $\mathcal{M}_{D+M}$
of Eq.~(\ref{D+M}).
The one-loop corrections generate an $n_L \times n_L$ submatrix $\delta M_L$
in the upper left-hand corner of $\mathcal{M}_{D+M}$,
where at tree level there is a zero submatrix.
The submatrices $M_D$ and $M_R$ also receive one-loop corrections,
denoted $\delta M_D$ and $\delta M_R$,
respectively.
We shall then have,
for the effective light-neutrino mass matrix at one loop,
\begin{equation}\label{M1}
\mathcal{M}_\nu = \mathcal{M}_\nu^\mathrm{tree} + \delta M_L -
\delta M_D^T M_R^{-1} M_D - M_D^T M_R^{-1} \delta M_D +
M_D^T M_R^{-1} \delta M_R M_R^{-1} M_D \,.
\end{equation}

One-loop corrections to the neutrino masses
originate in the one-loop neutrino self-energy $\Sigma(p)$,
where $p$ is the neutrino momentum.
We use the decomposition
\begin{equation}
\Sigma(p) = A_L \left( p^2 \right) \ppsl \gamma_L
+ A_R \left( p^2 \right) \ppsl \gamma_R
+ B_L \left( p^2 \right) \gamma_L
+ B_R \left( p^2 \right) \gamma_R \,.
\end{equation}
The non-absorptive parts of $A_{L,R}$ are Hermitian,
while those of $B_{L,R}$ are the Hermitian conjugates of each other.
Furthermore,
since the neutrino field vector $\chi$ consists of Majorana fields,
the self-energy must fulfil the consistency condition 
\begin{equation}
\Sigma(p) = C \left[ \Sigma(-p) \right]^T C^{-1} \,,
\end{equation}
and that consistency condition translates into
\begin{equation}
A_L = A_R^T \,, \quad B_L = B_L^T \,,
\quad \mathrm{and} \quad B_R = B_R^T \,.
\end{equation}

After computing the neutrino self-energies
one must renormalize the neutrino fields:
\begin{equation}
\chi_L = \left( 1 + \frac{1}{2}\, z_L \right) \chi_L^r\, ,
\quad \mathrm{hence} \quad
\chi_R = \left( 1 + \frac{1}{2}\, z_L^\ast \right) \chi_R^r\, ,
\end{equation}
since $\chi_R = C \bar \chi_L^T$.
The neutrino mass matrices must also be renormalized.
Mass counterterms are indicated by $\delta^c$.
The upper-left submatrix of $\mathcal{M}_{D+M}$,
which we call $M_L$,
has no counterterm since it vanishes at tree level:
$\delta^c M_L = 0$.
The operation $\delta^c$ only acts on $M_D$ and on $M_R$,
with,
from the second Eq.~(\ref{Mell,D}),
\begin{equation}\label{deltaMD}
\delta^c M_D = \frac{1}{\sqrt{2}} \sum_k \left[
\left( \delta^c v_k \right) \Delta_k
+ v_k \left( \delta^c \Delta_k \right) \right].
\end{equation}
Then,
\begin{equation}
\delta^c \hat m = \mathcal{U}^T
\left( \begin{array}{cc} 0 & \left( \delta^c M_D \right)^T \\
\delta^c M_D & \delta^c M_R
\end{array} \right) \mathcal{U}\, .
\end{equation}
The \emph{renormalized} neutrino self-energy,
in the basis where the tree-level neutrino mass matrix is diagonal,
has the structure \cite{denner,pilaftsis96,pilaftsis02} 
\begin{eqnarray}
-i \Sigma^r (p) &=& -i \Sigma (p)
+
i\, \frac{1}{2} \left( z_L + z_L^\dagger \right) \ppsl \gamma_L + 
i\, \frac{1}{2} \left( z_L^* + z_L^T \right) \ppsl \gamma_R 
\nonumber \\ && -
i \left( \delta^c \hat m + \frac{1}{2}\, \hat m z_L + 
\frac{1}{2}\, z_L^T \hat m \right) \gamma_L 
- i \left( \delta^c \hat m + \frac{1}{2}\, \hat m z_L + 
\frac{1}{2}\, z_L^T \hat m \right)^* \gamma_R \,. 
\label{sigmar}
\end{eqnarray}
We indicate one-loop contributions to $\Sigma (p)$ by $\dloop$.
There are tadpole diagrams indicated by $\delta^\mathrm{tadpole}$,
but we do not include them in $\Sigma (p)$.

Now we explain our strategy for calculating the dominant one-loop corrections
to the seesaw mechanism:
\begin{enumerate}
\renewcommand{\theenumi}{\roman{enumi}}
\item We calculate $\Sigma(p)$
in the basis where the tree-level neutrino mass matrix is diagonal.
\item We are especially interested in $B_L \left( p^2 \right)$,
because
\begin{equation}\label{dML}
\delta M_L = \dloop M_L = U_L^\ast B_L \left( 0 \right) U_L^\dagger \,.
\end{equation}
We have used the inverse of the relation (\ref{Mtotal})
to transform back to the basis of $\mathcal{M}_{D+M}$.
\item Denoting by $\Delta$ a typical coupling constant
of the coupling matrices $\Delta_k$,
the light-neutrino masses are of order $\Delta^2$---see Eq.~(\ref{Mtree}).
Therefore,
we are allowed to evaluate $B_L$ at $p^2 = 0$ in Eq.~(\ref{dML}).
Note that $\delta M_L$ is identical with $\dloop M_L$,
since $\delta^c M_L = 0$.
For the same reason,
$\dloop M_L$ must be finite.
We shall see this explicitly in the next section,
where we shall also demonstrate the gauge invariance of $\dloop M_L$.
\item On the other hand,
the one-loop corrections to $M_D$ and to $M_R$ are given by
\begin{equation}\label{dMD}
\dloop M_D = U_R\, B_L(0)\, U_L^\dagger
\quad \mathrm{and} \quad
\dloop M_R = U_R\, B_L(0)\, U_R^T \,,
\end{equation}
respectively.
We shall argue in Section \ref{general} that
one-loop corrections to $M_R$ are irrelevant,
while those to $M_D$ lead to subdominant corrections in $\mathcal{M}_\nu$. 
\end{enumerate}

First we consider the contribution of the one-loop graph with $W^\pm$
to $\Sigma (p)$:
\begin{eqnarray}
\Sigma^{(W)}_{ij} (p) &=& i\, \frac{g^2}{2}
\int \! \frac{d^d k}{\left( 2 \pi \right)^d}
\left[ S^W_{\mu \nu} \left( k-p \right) \right]
\sum_\ell \frac{1}{k^2 - m_\ell^2}
\nonumber\\ & &
\times \left[
\left( U_L^\dagger \right)_{i\ell} \left( U_L \right)_{\ell j}
\gamma^\mu \kksl \gamma^\nu \gamma_L
+ \left( U_L^T \right)_{i\ell} \left( U_L^\ast \right)_{\ell j}
\gamma^\mu \kksl \gamma^\nu \gamma_R
\right],
\end{eqnarray}
where $i S^W_{\mu \nu} \left( k-p \right)$
is the propagator of a $W^\pm$ with momentum $k-p$,
and $d$ is the dimension of space--time.
Clearly,
$\Sigma^{(W)}_{ij} (p)$ only contributes to $A_L \left( p^2 \right)$
and $A_R \left( p^2 \right)$.

Next we consider the contribution from the $Z$ boson.
This yields,
in a general $R_\xi$ gauge parameterized by the gauge parameter $\xi_Z$,
\begin{eqnarray}
\left( B_L \right)_{ij}^{(Z)} \left( p^2 \right) &=&
i\, \frac{g^2}{4 c_w^2}\, \int \! \frac{d^d k}{\left( 2 \pi \right)^d}
\sum_\iota \frac{m_\iota}{k^2 - m_\iota^2}
\left( U_L^T U_L^\ast \right)_{i\iota}
\left( U_L^\dagger U_L \right)_{\iota j}
\nonumber\\ & &
\times
\left\{ \frac{d}{\left( k-p \right)^2 - m_Z^2}
+ \frac{1}{m_Z^2}
\left[ \frac{\left( k-p \right)^2}{\left( k-p \right)^2 - \xi_Z m_Z^2}
- \frac{\left( k-p \right)^2}{\left( k-p \right)^2 - m_Z^2} \right] \right\}.
\hspace*{5mm}
\label{Z}
\end{eqnarray}

Now we consider the contribution of the one-loop graph with $S_a^\pm$.
It is
\begin{eqnarray}
\left( B_L \right)_{ij}^{(S^\pm)} \left( p^2 \right) &=&
-i \int \! \frac{d^d k}{\left( 2 \pi \right)^d}
\sum_a \frac{1}{\left( k-p \right)^2 - m_a^2}
\sum_\ell \frac{m_\ell}{k^2 - m_\ell^2}
\nonumber\\ & &
\times \left[ \left( U_R^\dagger \Delta_a \right)_{i\ell}
\left( \Gamma_a U_L \right)_{\ell j}
+ \left( U_L^T \Gamma_a^T \right)_{i\ell}
\left( \Delta_a^T U_R^\ast \right)_{\ell j}
\right].
\label{S+-}
\end{eqnarray}
When $S_a^\pm$ is the charged Goldstone boson $G^\pm$,
one uses Eqs.~(\ref{ugjsl}) together with Eq.~(\ref{relation}) to derive
\begin{eqnarray}
\left( B_L \right)_{ij}^{(G^\pm)} \left( p^2 \right) &=&
-i\, \frac{g^2}{2 m_W^2}\, \int \! \frac{d^d k}{\left( 2 \pi \right)^d}\,
\frac{1}{\left( k-p \right)^2 - \xi_W m_W^2}
\sum_\ell \frac{m_\ell^2}{k^2 - m_\ell^2}
\nonumber\\ & &
\times \left[ m_i \left( U_L^\dagger \right)_{i\ell}
\left( U_L \right)_{\ell j}
+ \left( U_L^T \right)_{i\ell}
\left( U_L^\ast \right)_{\ell j} m_j
\right].
\end{eqnarray}

In similar fashion,
the neutral scalars $S^0_b$ contribute
(neglecting tadpoles)
\begin{eqnarray}
\left( B_L \right)_{ij}^{(S^0)} \left( p^2 \right) &=&
\frac{i}{2}\, \int \! \frac{d^d k}{\left( 2 \pi \right)^d}
\sum_b \frac{1}{\left( k-p \right)^2 - m_b^2}
\sum_\iota \frac{m_\iota}{k^2 - m_\iota^2}
\nonumber\\ & &
\times \left( U_R^\dagger \Delta_b U_L
+ U_L^T \Delta_b^T U_R^\ast \right)_{i\iota}
\left( U_R^\dagger \Delta_b U_L
+ U_L^T \Delta_b^T U_R^\ast \right)_{\iota j}\, .
\label{S0}
\end{eqnarray}
If the scalar $S^0_b$ is the neutral Goldstone boson $G^0$,
then,
using Eqs.~(\ref{utiwm}) and (\ref{relation}),
\begin{eqnarray}
\left( B_L \right)_{ij}^{(G^0)} \left( p^2 \right) &=&
- \frac{i g^2}{4 m_W^2}\,
\int \! \frac{d^d k}{\left( 2 \pi \right)^d}\,
\frac{1}{\left( k-p \right)^2 - \xi_Z m_Z^2}
\sum_\iota \frac{m_\iota}{k^2 - m_\iota^2}
\nonumber\\ & &
\times \left( \hat m U_L^\dagger U_L
+ U_L^T U_L^\ast \hat m \right)_{i\iota}
\left( \hat m U_L^\dagger U_L
+ U_L^T U_L^\ast \hat m \right)_{\iota j}\, .
\end{eqnarray}

\section{The computation of $\delta M_L$} \label{deltaML}

In this section,
we explicitly perform the calculation of $\delta M_L$,
defined in Eq.~(\ref{dML}). 

It is clear from Eq.~(\ref{S+-}) and from the third Eq.~(\ref{uni})
that the exchange of $S^\pm$ contributes neither to $\dloop M_L$
nor to $\dloop M_R$;
it only contributes to $\dloop M_D$.
We therefore have
\begin{equation}
\delta M_L = \delta M_L \left( Z \right) + \delta M_L \left( G^0 \right)
+ \sum_{b \neq b_Z} \delta M_L \left( S^0_b \right).
\end{equation}
The $Z$ boson part is given by
\begin{equation}
\delta M_L \left( Z \right) =
i\, \frac{g^2}{4 c_w^2}\, \int \! \frac{d^d k}{\left( 2 \pi \right)^d}\,
\left[ \frac{d}{k^2 - m_Z^2}
+ \frac{1}{m_Z^2}
\left( \frac{k^2}{k^2 - \xi_Z m_Z^2} - \frac{k^2}{k^2 - m_Z^2} \right)
\right]
U_L^\ast \frac{\hat m}{k^2 - \hat m^2} U_L^\dagger \,.
\label{ubdoq}
\end{equation}
The neutral Goldstone boson contributes
\begin{equation}
\delta M_L \left( G^0 \right) =
- \frac{i g^2}{4 m_W^2}\,
\int \! \frac{d^d k}{\left( 2 \pi \right)^d}\, \frac{1}{k^2 - \xi_Z m_Z^2}\,
U_L^\ast \frac{\hat m^3}{k^2 - \hat m^2} U_L^\dagger \, .
\label{kfirt}
\end{equation}
Using $m_W^2 = c_w^2 m_Z^2$ and the relation
\begin{equation}
U_L^\ast\, \frac{k^2 \hat m}{k^2 - {\hat m}^2}\, U_L^\dagger =
U_L^\ast\, \frac{{\hat m}^3}{k^2 - {\hat m}^2}\, U_L^\dagger \,,
\end{equation}
which holds by virtue of $U_L^\ast \hat m U_L^\dagger = 0$,
it is clear that the $\xi_Z$-dependent terms
in Eqs.~(\ref{ubdoq}) and (\ref{kfirt}) cancel
and therefore \emph{$\delta M_L$ is gauge-invariant}.
One has
\begin{equation}
\delta M_L \left( Z \right) + \delta M_L \left( G^0 \right) =
i\, \frac{g^2}{4 c_w^2}\, \int \! \frac{d^d k}{\left( 2 \pi \right)^d}\,
\frac{1}{k^2 - m_Z^2}\,
U_L^\ast \left( d - \frac{\hat m^2}{m_Z^2} \right)
\frac{\hat m}{k^2 - \hat m^2}\, U_L^\dagger\, .
\label{jclwo}
\end{equation}
The contribution to $\delta M_L$ from the exchange of all neutral scalars,
except the neutral Goldstone boson,
is
\begin{equation}
\delta M_L \left( S^0_b \right) =
\frac{i}{2}\, \int \! \frac{d^d k}{\left( 2 \pi \right)^d}\,
\frac{1}{k^2 - m_b^2}\,
\Delta_b^T U_R^\ast\, \frac{\hat m}{k^2 - \hat m^2}\, U_R^\dagger \Delta_b\, .
\end{equation}

Now we perform the actual calculation of the integrals.
Let the dimension of space--time be $d = 4 - 2 \epsilon$,
with $\epsilon \to 0$.
We define the divergent quantity
\begin{equation}
k = - \epsilon^{-1} + \gamma - \ln \left( 4 \pi \right) - 1\, ,
\end{equation}
where $\gamma$ is Euler's constant.
Then,
we obtain
\begin{equation}\label{Sf}
\delta M_L \left( S_b^0 \right) =
\frac{1}{32 \pi^2}\, \Delta_b^T U_R^\ast \hat m
\left( k + \ln{\hat m^2} + \frac{\ln{r_b}}{r_b - 1} \right)
U_R^\dagger \Delta_b\, ,
\quad \mathrm{with} \quad
r_b = \frac{\hat m^2}{m_b^2}\, .
\end{equation}
Similarly,
\begin{equation}\label{Z1f}
\delta M_L \left( Z, 1 \right) =
\frac{g^2}{16 \pi^2 c_w^2}\, U_L^\ast \hat m
\left( k + \frac{1}{2} + \ln{\hat m^2} + 
\frac{\ln{r_Z}}{r_Z-1} \right) U_L^\dagger\, ,
\quad \mathrm{with} \quad
r_Z = \frac{\hat m^2}{m_Z^2}\, ,
\end{equation}
is the contribution
to $\delta M_L \left( Z \right) + \delta M_L \left( G^0 \right)$
from the term with $d$ in Eq.~(\ref{jclwo});
the term with $- \hat m^2 / m_Z^2$ yields
\begin{equation}\label{Z2f}
\delta M_L \left( Z, 2 \right) =
- \frac{g^2}{64 \pi^2 m_W^2}\, U_L^\ast \hat m^3
\left( k + \ln{\hat m^2} + 
\frac{\ln{r_Z}}{r_Z-1} \right) U_L^\dagger\, .
\end{equation}

In order to demonstrate the cancellation of infinities
we need the orthogonality relation
of the vectors $b \in \mathbbm{C}^{n_H}$ \cite{GN89}:
\begin{equation}\label{sumb}
\sum_{b \neq b_Z} b_j b_k
+ \left( b_Z \right)_j \left( b_Z \right)_k = 0\, .
\end{equation}
Remembering that $\Delta_b = \sum_k b_k \Delta_k$,
we find that all terms independent of the boson masses---in particular
the infinities---drop out in the sum $\sum_{b \neq b_Z}
\delta M_L \left( S_b^0 \right) + \delta M_L \left( Z, 2 \right)$.
We may thus write
\begin{eqnarray}
\delta M'_L \left( S_b^0 \right) &=&
\frac{1}{32 \pi^2}\, \Delta_b^T U_R^\ast \hat m\,
\frac{\ln{r_b}}{r_b - 1}\, U_R^\dagger \Delta_b\, ,
\\
\delta M'_L \left( Z, 2 \right) &=&
- \frac{g^2}{64 \pi^2 m_W^2}\, U_L^\ast {\hat m}^3\,
\frac{\ln{r_Z}}{r_Z - 1}\, U_L^\dagger \,.
\end{eqnarray}
Also,
because of the first Eq.~(\ref{0}),
the terms in $\delta M_L \left( Z, 1 \right)$
of Eq.~(\ref{Z1f})
which are proportional to $\hat m$
(but not the term with $\hat m \ln{\hat m^2}$),
in particular the infinities,
drop out.
Thus,
\begin{equation}
\delta M'_L \left( Z, 1 \right) =
\frac{g^2}{16 \pi^2 c_w^2}\, U_L^\ast \hat m\, 
\frac{r_Z \ln{r_Z}}{r_Z - 1}\, U_L^\dagger\, .
\end{equation}
It is obvious that
$\delta M'_L \left( Z, 1 \right)
= - 4\, \delta M'_L \left( Z, 2 \right)$
We thus we have the final result
\begin{equation}\label{final}
\delta M_L =
\sum_{b \neq b_Z} \frac{1}{32 \pi^2}\, \Delta_b^T U_R^\ast \hat m\,
\frac{\ln{r_b}}{r_b - 1} U_R^\dagger \Delta_b
+ \frac{3 g^2}{64 \pi^2 m_W^2}\, U_L^\ast \hat m^3\,
\frac{\ln{r_Z}}{r_Z - 1}\, U_L^\dagger\, .
\end{equation}
Using Eq.~(\ref{relation}),
the last term of the previous relation reads
\begin{equation}
\frac{3 g^2}{64 \pi^2 m_W^2}\, M_D^T U_R^\ast \hat m\,
\frac{\ln{r_Z}}{r_Z - 1}\, U_R^\dagger M_D\,.
\end{equation}
This expression,
together with $m_W = g v / 2$, 
shows that the $Z$ contribution to $\delta M_L$
is of the same order of magnitude as the neutral-scalar contributions.

\section{Dominance of $\delta M_L$ among the radiative corrections}
\label{general}

We now want to give arguments for the dominance of $\delta M_L$
in the radiative corrections to $\mathcal{M}_\nu^\mathrm{tree}$.
The main point in our argumentation is the observation that
the terms in $\delta M_L$ in Eq.~(\ref{final})
are smaller than the tree-level masses of the light neutrinos
solely because of the factor $\left( 16\pi^2 \right)^{-1}$
from the one-loop integrals.
In this context,
we have to remember that,
in our framework,
it is natural to assume all scalar masses
to be of the order of the electroweak scale. 

The heavy-neutrino masses $m_{n_L+1}, \dots, m_{n_L+n_R}$
(or the elements of $M_R$)
are free parameters of the theory;
therefore,
corrections to $M_R$ are irrelevant for us. 
We stress that this contrasts to the light-neutrino masses
$m_1, m_2, \dots, m_{n_L}$,
which are calculable,
i.e., 
they are functions of the parameters of the theory;
this is reflected in the finiteness and gauge-invariance of $\delta M_L$.

Concentrating on $\dloop M_D$ as given by Eq.~(\ref{dMD}), 
we note that both $S^\pm_a$ and $S_b^0$ contribute to it,
see Eqs.~(\ref{S+-}) and (\ref{S0}),
respectively.
Checking Eq.~(\ref{Z}),
we note the absence of a $Z$ contribution.
In the integral of Eq.~(\ref{S+-}),
for charged-scalar exchange,
the large mass scale $m_R$ is absent;
therefore,
if $Y$ is a typical Yukawa coupling,
then we must have $\dloop M_D(S_a^\pm) \sim Y^2 m_\ell$,
where $m_\ell$ is a generic charged-lepton mass.
With $m_\ell \sim m_D$,
we have $Y^2 m_D^2 / m_R$ as the order of magnitude
of the charged-scalar contribution to $\mathcal{M}_\nu$,
which is suppressed as compared to the $\delta M_L$ contribution.
Turning now to $S_b^0$ exchange,
with Eqs.~(\ref{dMD}) and (\ref{S0}) the result of the one-loop calculation is 
\begin{equation}
\dloop M_D \left( S_b^0 \right) = \frac{1}{32\pi^2}\, 
\Delta_b U_L \hat m \left( k + \ln{\hat m^2} + \frac{\ln r_b}{r_b - 1} 
\right) U_R^\dagger \Delta_b\,.
\end{equation}
Again,
we arrive at $\dloop M_D(S_b^0) \sim Y^2 m_D$.
Therefore,
the contributions to $\mathcal{M}_\nu$ of Eq.~(\ref{M1})
from $\dloop M_D$ are negligible when compared
to the ones from $\delta M_L$. 

Some remarks concerning the wave function renormalization matrix $z_L$
are in order.
Usually,
that matrix is defined
in such a way that the renormalized fields remain in the mass basis
\cite{denner,pilaftsis96,pilaftsis02}.
With Eqs.~(\ref{Mtree}) and (\ref{dML}) we do not adopt this convention:
the mass matrix of the light neutrinos,
$\mathcal{M}_\nu$,
is not diagonal at order $Y^2$.
This is consistent with the fact that,
at zeroth order in the Yukawa couplings,
the light neutrinos are massless and,
therefore,
degenerate.
This degeneracy is lifted only at order $Y^2$,
and a large rotation,
with angles of order one,
will be needed,
in general,
for the diagonalization of $\mathcal{M}_\nu$.
Since we assume that $z_L$ is ``small'' as compared to one,
we do not include the large rotation in this matrix.
In the same vein, 
concerning the terms $\hat m z_L + z_L^T \hat m$ in Eq.~(\ref{sigmar}),
we are allowed to set to zero the light-neutrino masses.
Using the explicit form of $z_L$ for on-shell renormalization
of the \emph{heavy} neutrinos,
as found,
for instance,
in Refs.~\cite{GL02,pilaftsis96},
and taking into account that
the small elements of $\mathcal{U}$ are of order $m_D/m_R$,
we find that the wave function renormalization terms of Eq.~(\ref{sigmar})
contribute at order $Y^2 m_D^2 / m_R$ or $g^2 m_D^2 / m_R$
to $\mathcal{M}_\nu$,
and then we do not need to consider them further.

The tadpole graphs contribute only to $\delta M_D$.
For their treatment
and their relation to the $\delta^c v_k$ terms in Eq.~(\ref{deltaMD}),
see the extensive discussion in Ref.~\cite{weinberg}.
They have no impact on our discussion.

\section{Conclusions} \label{concl}

We have argued in the preceding section that
the dominant radiative corrections to $\mathcal{M}_\nu^\mathrm{tree}$
are given by Eq.~(\ref{dML}),
which has contributions only from neutral-scalar and $Z$ exchange;
equation~(\ref{final}) is the explicit form of Eq.~(\ref{dML})
after carrying out the one-loop calculations. 
These radiative corrections  are,
like the tree-level mass term of Eq.~(\ref{Mtree}),
quadratic in the Yukawa couplings. 
In Eq.~(\ref{final})---as compared
to $\mathcal{M}^\mathrm{tree}_\nu$---only
the terms of first order in $1/m_R$ are relevant
($m_R$ is the scale of $M_R$).
We may use the approximations 
\begin{equation}
U_R \simeq \left( 0, W \right),
\quad \mathrm{with} \quad
W^\dagger M_R W^\ast \simeq \widetilde m \equiv
\mathrm{diag}\, (m_{n_L+1}, \ldots, m_{n_L+n_R}) \,,
\end{equation}
where $W$ is a unitary $n_R \times n_R$ matrix
whose elements are not suppressed by $m_D/m_R$
($m_D$ is the scale of $M_D$). 
We thus have the final result
\begin{equation}\label{mm}
\mathcal{M}_\nu = - M_D^T M_R^{-1} M_D + \delta M_L\, ,
\end{equation}
with
\begin{eqnarray}
\delta M_L & = & \nonumber
\sum_{b \neq b_Z} \frac{m_b^2}{32\pi^2}\, \Delta_b^T W^\ast
\left( \frac{1}{\widetilde m} \ln{\frac{\widetilde m^2}{m_b^2}} \right) 
W^\dagger \Delta_b \\
&& + \frac{3 g^2}{64 \pi^2 c_w^2}\, M_D^T W^\ast
\left( \frac{1}{\widetilde m} \ln{\frac{\widetilde m^2}{m_Z^2}} \right) 
W^\dagger M_D \,.
\label{final1}
\end{eqnarray}
The sum over the neutral scalars includes only the physical ones.
The second line in Eq.~(\ref{final1}) is the $Z$ contribution.
Note that if accidentally
$\mathcal{M}_\nu^\mathrm{tree} =
- M_D^T W^\ast \left( 1 / \tilde m \right) W^\dagger M_D$ vanishes,
then, even for a single Higgs doublet,
the one-loop $Z$ and Higgs contributions to
$\delta M_L$ will in general be non-zero \cite{pilaftsis92}.

Having more than one Higgs doublet leads to flavour-changing neutral
interactions.
Neutral-scalar exchange at tree level
then leads to processes like $\mu^\pm \to e^\pm e^+ e^-$; 
however, such processes only restrict
elements of the coupling matrices $\Gamma_k$, which do not occur in
$\mathcal{M}_\nu$ of Eq.~(\ref{mm}).
Moreover,
with more than one Higgs doublet and at one-loop level, radiative
decays like $\mu^\pm \to e^\pm \gamma$ proceed via charged-scalar
exchange. However, moderately large charged-scalar masses, still in the range
of the electroweak scale, can render their decay rates compatible
with experimental bounds \cite{GN00}.

Working in the physical basis of the charged leptons,
Eqs.~(\ref{mm}) and (\ref{final1}) determine the neutrino masses and mixings
in the Standard Model with an arbitrary number of Higgs doublets.
As we have shown,
$\delta M_L$ is finite and gauge-invariant.
Moreover,
we have argued that all other one-loop contributions to $\mathcal{M}_\nu$
are of order $Y^2 m_D^2 / m_R$ (or $g^2 m_D^2 / m_R$),
where $Y$ is a typical Yukawa coupling constant;
consequently,
$\delta M_L$ dominates the one-loop corrections.

The fact that the order of magnitude of the one-loop corrections
in Eq.~(\ref{final1}) is smaller than the tree-level mass term
solely by a factor of order
$\left( 16 \pi^2 \right)^{-1} \ln \left( m_R / m_0 \right)$,
where $m_0$ is a typical neutral-scalar mass,
can be exploited for generating
the ratio $\Delta m^2_\odot/\Delta m^2_\mathrm{atm}$ radiatively,
such that the order of magnitude of that ratio
is precisely the one adequate for the large-mixing-angle MSW solution
of the solar-neutrino problem \cite{GN00}.
However,
the question of whether this framework also allows
to explain the specific features of lepton mixing,
namely the small element $U_{e3}$ of the mixing matrix,
and an atmospheric mixing angle close to $45^\circ$,
requires further investigation.

\end{document}